\documentclass{article}
\usepackage{spconf,amsmath,graphicx}
\usepackage[numbers,compress]{natbib}
\usepackage{makecell}
\usepackage{tabularx}
\usepackage{amssymb}

\usepackage[utf8]{inputenc} 
\usepackage[T1]{fontenc}    
\usepackage{hyperref}       
\usepackage{url}            
\usepackage{booktabs}       
\usepackage{amsfonts}       
\usepackage{nicefrac}       
\usepackage{microtype}      
\usepackage{xcolor}         
\usepackage{multirow}
\usepackage{algorithmic}
\usepackage{array}
\usepackage{textcomp}
\usepackage{stfloats}
\usepackage{verbatim}
\usepackage{subcaption}
\usepackage{tikz}
\usepackage{pgf-pie}
\usepackage{mathptmx}       

\makeatletter
\newcommand\footnoteref[1]{\protected@xdef\@thefnmark{\ref{#1}}\@footnotemark}
\makeatother

\def\BibTeX{{\rm B\kern-.05em{\sc i\kern-.025em b}\kern-.08em T\kern-.1667em\lower.7ex\hbox{E}\kern-.125emX}}

\usepackage{times}
\usepackage{latexsym}
\usepackage{url}
\usepackage[utf8]{inputenc}
\usepackage{microtype}
\usepackage{inconsolata}
\usepackage{enumitem}
\usepackage{pifont}
\usepackage{placeins}

\usetikzlibrary{shapes.geometric, arrows}
\tikzstyle{startstop} = [rectangle, rounded corners, minimum width=3cm, minimum height=1cm, text centered, draw=black]
\tikzstyle{process} = [rectangle, minimum width=3cm, minimum height=1cm, text centered, draw=black]
\tikzstyle{arrow} = [thick,->,>=stealth]

\usepackage{listings}
\usepackage{xcolor}
\usepackage{newtxtext,newtxmath}

\definecolor{entitytag}{HTML}{0550ae}
\definecolor{keyword}{HTML}{d73a49}
\definecolor{value}{HTML}{22863a}
\definecolor{string}{HTML}{032f62}
\definecolor{background}{HTML}{f6f8fa}

\lstdefinelanguage{json}{
    basicstyle=\normalfont\ttfamily,
    numbers=left,
    numberstyle=\tiny,
    stepnumber=1,
    numbersep=8pt,
    showstringspaces=false,
    breaklines=true,
    frame=lines,
    backgroundcolor=\color{background},
    literate=
     *{0}{{{\color{value}0}}}{1}
      {1}{{{\color{value}1}}}{1}
      {2}{{{\color{value}2}}}{1}
      {3}{{{\color{value}3}}}{1}
      {4}{{{\color{value}4}}}{1}
      {5}{{{\color{value}5}}}{1}
      {6}{{{\color{value}6}}}{1}
      {7}{{{\color{value}7}}}{1}
      {8}{{{\color{value}8}}}{1}
      {9}{{{\color{value}9}}}{1}
      {:}{{{\color{keyword}:}}}{1}
      {,}{{{\color{keyword},}}}{1}
      {"}{{{\color{string}"}}}{1}
      {true}{{{\color{value}true}}}{4}
      {false}{{{\color{value}false}}}{5}
      {null}{{{\color{value}null}}}{4}
      {\{}{{{\color{entitytag}\{}}}{1}
      {\}}{{{\color{entitytag}\}}}}{1}
      {\ [}{{{\color{entitytag}[}}}{1}
      {\ ]}{{{\color{entitytag}]}}}{1}
}

\title{Emilia: An Extensive, Multilingual, and Diverse Speech Dataset for Large-Scale Speech Generation}

\name{
\begin{tabular}{c}
  Haorui He$^{1,\star}$\thanks{$^{\star}$ Equal contribution, and the names are listed in random order.} \qquad  
  Zengqiang Shang$^{2,\star}$ \qquad 
  Chaoren Wang$^{1,\star}$ \qquad 
  Xuyuan Li$^{2,3,\star}$ \qquad \\
  Yicheng Gu$^{1}$ \qquad 
  Hua Hua$^{2,3}$ \qquad 
  Liwei Liu$^{1}$ \qquad 
  Chen Yang$^{2,3}$ \qquad 
  Jiaqi Li$^{1}$ \qquad 
  Peiyang Shi$^{2}$ \qquad \\
  Yuancheng Wang$^{1}$ \qquad
  Kai Chen$^{4}$ \qquad 
  Pengyuan Zhang$^{2,3,\ddagger}$ \qquad 
  Zhizheng Wu$^{1,4,\ddagger}$\thanks{$^{\ddagger}$ Corresponding authors.}  
\end{tabular}
}

\address{
\begin{tabular}{c}
  $^{1}$ The Chinese University of Hong Kong, Shenzhen, China\\
  $^{2}$ Laboratory of Speech \& Intelligent Information Processing, Institute of Acoustics, CAS, China\\
  $^{3}$ University of Chinese Academy of Sciences, Beijing, China\\
  $^{4}$ Shanghai AI Laboratory, Shanghai, China
\end{tabular}
}

\begin{document}
\ninept
\maketitle

\begin{abstract}
Recent advancements in speech generation models have been significantly driven by the use of large-scale training data. However, producing highly spontaneous, human-like speech remains a challenge due to the scarcity of large, diverse, and spontaneous speech datasets. In response, we introduce \textit{Emilia}, the first large-scale, multilingual, and diverse speech generation dataset. Emilia starts with over 101k hours of speech across six languages, covering a wide range of speaking styles to enable more natural and spontaneous speech generation. To facilitate the scale-up of Emilia, we also present \textit{Emilia-Pipe}, the first open-source preprocessing pipeline designed to efficiently transform raw, in-the-wild speech data into high-quality training data with speech annotations. Experimental results demonstrate the effectiveness of both Emilia and Emilia-Pipe. Demos are available at: \url{https://emilia-dataset.github.io/Emilia-Demo-Page/}.
\end{abstract}


\section{Introduction}
In the past few years, the research of speech generation has made significant advancements with the emergence of various generative models~\cite{borsos2023soundstorm, voicebox, ju2024naturalspeech, gyc} and the use of large-scale training data~\cite{librilight, mls}. The models such as SoundStorm~\cite{borsos2023soundstorm}, VoiceBox~\cite{le2024voicebox} and NaturalSpeech 3~\cite{ju2024naturalspeech} have considerably progressed in (zero-shot) speech generation by considerably scaling up both the datasets and model sizes, achieving high similarity, voice quality, and naturalness on academic datasets~\cite{amphion}. However, the generated speech still fails to generate highly spontaneous and human-like speech in the real world~\cite{ju2024naturalspeech,tan2021survey}.

One of the significant reasons for such limitation is that the current speech generation models are trained on speech datasets such as Libri-Light \cite{librilight} and MLS \cite{mls}, which have their root in audiobooks. Such datasets typically are characterized by formal reading styles. However, speech from real humans, especially in casual or conversational contexts, rarely adheres to such standardized patterns. Instead, it exhibits more diverse and spontaneous speaking styles, including breathing, pausing, repetitions, changes in speed, and varying emotions. Consequently, there is a pressing need for a new dataset that encompasses more diverse speech styles to advance the field towards generating more spontaneous and human-like speech. 

However, directly using in-the-wild speech data is not feasible due to variations in length and quality, frequent background noise, music, reverberation, the presence of multiple speakers within a single sample, and the lack of necessary annotations such as text transcriptions~\cite{AutoPrep}. Training with such data may degrade the performance of speech generation models. 
While previous works~\cite{AutoPrep,wenetspeech4tts} propose automatic preprocessing pipelines to address such issues, they depend heavily on proprietary models, limiting their accessibility to the wider community.
Additionally, the processing speed of these pipelines remains unknown. An ideal preprocessing pipeline for in-the-wild speech data should operate quickly to handle large volumes efficiently, enabling significant dataset scaling.  Furthermore, the resulting datasets from these pipelines are limited to monolingual (Chinese-only) data and are relatively small in size (39 hours for~\cite{AutoPrep}, 12k hours for~\cite{wenetspeech4tts}).

In response, this paper introduces \textit{Emilia-Pipe}, the first open-source preprocessing pipeline designed to efficiently convert in-the-wild speech data into high-quality, annotated training data for speech generation. \textit{Emilia-Pipe} integrates several engineering optimizations to enhance both robustness and efficiency. The pipeline supports multiple languages and processes one hour of raw speech data in just a few minutes, enabling broader collaboration within the research community on large-scale speech generation.
Leveraging Emilia-Pipe, we construct the first multilingual speech generation dataset from in-the-wild speech data, \textit{Emilia}. Table~\ref{tab:speech_datasets} compares Emilia with several existing speech generation datasets. The key advantages of the Emilia dataset are summarized as follows:

\begin{itemize}
\item \textbf{Extensive and Multilingual.} The Emilia dataset contains over 101k hours of speech data at 24 kHz and covers six languages: English (En), Chinese (Zh), German (De), French (Fr), Japanese (Ja), and Korean (Ko). To the best of our knowledge, it is the largest academic speech generation dataset.
\item \textbf{Diverse.} The Emilia dataset comprises mostly spontaneous speech, covering a wide range of speaking styles. This diversity is crucial for training high-quality models to generate spontaneous and human-like speech generation models.
\item \textbf{Dynamic.} The Emilia dataset features an automatic and efficient preprocessing pipeline, Emilia-Pipe, which allows for easy scaling by incorporating readily available raw in-the-wild speech data or other user-specified source speech data.
\end{itemize}

To validate the effectiveness of \textit{Emilia}, two text-to-speech (TTS) models are trained on the English subset of the Emilia dataset and compared with their counterparts trained on the MLS audiobook dataset. Experimental results from both subjective and objective evaluations demonstrate that \textit{Emilia} is highly effective for training models capable of generating high-quality, spontaneous, and human-like speech. Furthermore, models trained on the full \textit{Emilia} dataset show promising performance in multilingual TTS tasks.

The Emilia-Pipe and Emilia dataset, is now publicly available.\footnote{GitHub: \url{https://github.com/open-mmlab/Amphion/tree/main/preprocessors/Emilia}; HuggingFace: \url{https://huggingface.co/datasets/amphion/Emilia-Dataset}.}

\begin{table*}[htbp]
    \centering
    \caption{A comparison of Emilia with existing datasets for speech generation. Note that the pipeline in \cite{AutoPrep} and \cite{wenetspeech4tts} is not publicly-available.}
    \label{tab:speech_datasets}
    \resizebox{\textwidth}{!}{
        \begin{tabular}{ccccccc}
            \toprule
            \textbf{Dataset} & \textbf{Data Source} & \textbf{Total Duration (hours)} & \textbf{Lang.} & \textbf{Samp. Rate (Hz)}  & \textbf{Dynamic} \\
            \midrule
            LJSpeech~\cite{ljspeech} & Audiobook & 24 & En & 22.05k  &   \\
            AutoPrepWild~\cite{AutoPrep} & In-the-wild & 39 & Zh & 24k/44.1k   & \checkmark (not open-source)\\
            VCTK~\cite{vctk} & Studio Recording & 44 & En & 48k  &  \\
            Aishell-3~\cite{aishell3}& Studio Recording  & 85 & Zh & 44.1k   &   \\
            LibriTTS~\cite{libritts} & Audiobook & 585 & En & 24k & &  \\
            GigaSpeech~\cite{gigaspeech}& In-the-wild & 10k & En & 16k &   \\
            WenetSpeech4TTS~\cite{wenetspeech4tts}& In-the-wild & 12k & Zh & 16k  & \checkmark (not open-source) \\
            MLS~\cite{mls} & Audiobook & 51k & En/Fr/De/Nl/Es/It/Pt/Pl & 16k &  \\ 
            Libri-Light~\cite{librilight} & Audiobook & 60k & En & 16k  &  \\
            \midrule
            Emilia & In-the-wild & 101k  & En/Zh/De/Fr/Ja/Ko& 24k   & \checkmark \\
            \bottomrule
        \end{tabular}
    }
\end{table*}

\section{The Emilia-Pipe Preprocessing Pipeline} \label{sec:pipline}
\begin{figure*}[ht]
    \centering
    \includegraphics[width=0.69\textwidth]{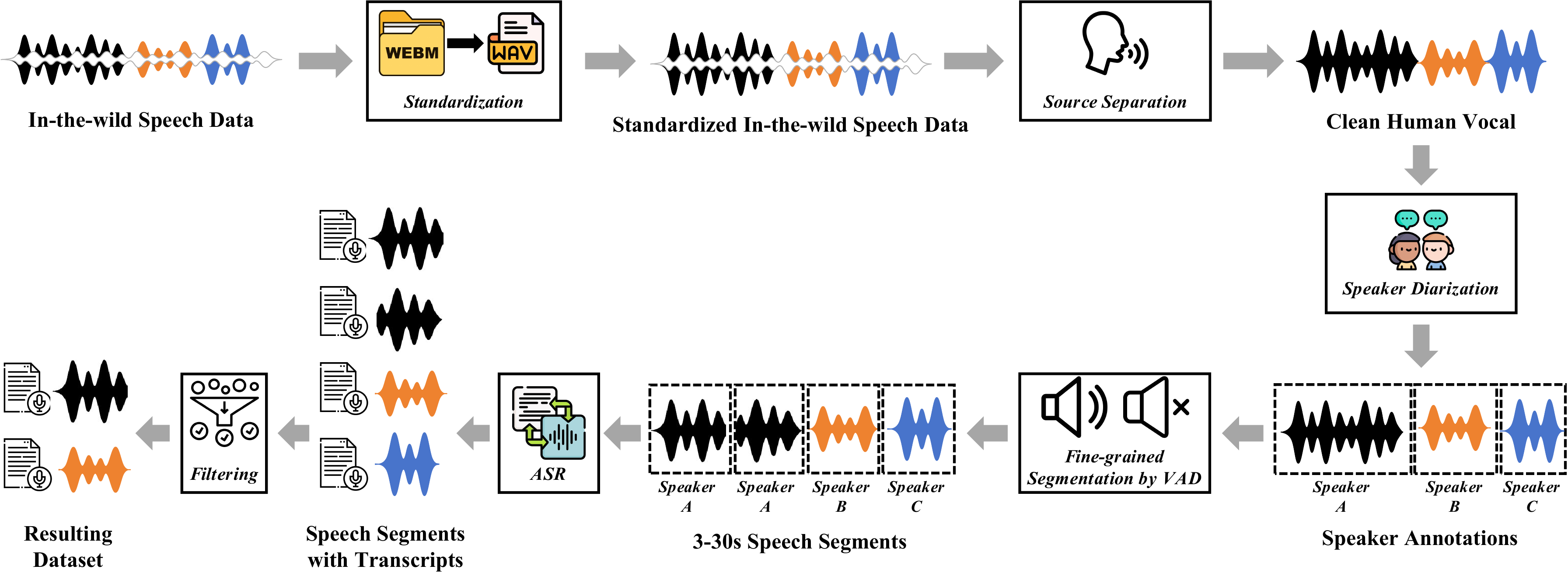}
    \caption{An overview of the Emilia-Pipe preprocessing pipeline.}
    \label{fig:process_pipeline}
\end{figure*}

This section details Emilia's data preprocessing pipeline, Emilia-Pipe. As illustrated in Fig.~\ref{fig:process_pipeline}, Emilia-Pipe includes six steps, i.e., Standardization, Source Separation, Speaker Diarization, Fine-grained Segmentation by VAD, ASR, and Filtering.

\subsection{Standardization}
The raw speech data in the wild vary in encoding formats, sampling rates, etc. Therefore we convert all samples to WAV format, set them to a mono channel, and resample to 24 kHz. We set the sample width to 16-bit and adjust the target decibel level to -20 dBFS. The volume adjustment is constrained between -3 and 3 dB to avoid distortion.
Finally, we normalize the waveform by dividing each sample by the maximum amplitude, ensuring values range between -1 and 1. These steps ensure a consistent data format for further processing.

\subsection{Source Separation}
The raw speech data in the wild often contain background music/noise, which negatively impacts speech generation performance~\cite{voicebox,how_far_robust_vc}. To address this, we use the source separation technique to extract clean human vocals. Specifically, we utilize the Ultimate Vocal Remover,\footnote{\url{https://github.com/Anjok07/ultimatevocalremovergui}} and its pre-trained model, UVR-MDX-Net Inst 3\footnote{\url{https://github.com/TRvlvr/model_repo/releases/tag/all_public_uvr_models}}, to separate human vocals from raw speech data for further processing.

\subsection{Speaker Diarization}
After extracting clean human vocals from the raw speech data, we apply the speaker diarization technique to partition the long-form speech data into multiple segments based on the speaker. This process generates a series of segments for each speech data, with each utterance containing only one speaker, ensuring compatibility with existing datasets for speech generation~\cite{librilight,mls,vctk,libritts}. To achieve this, we leverage the ``pyannote/speaker-diarization-3.1'' speaker diarization pipeline.\footnote{\url{https://github.com/pyannote/pyannote-audio}} This pipeline includes three core components: speaker segmentation, speaker embedding, and clustering, and achieves state-of-the-art speaker diarization performance~\cite{pyannote_model}. 
The output of this pipeline is a list of temporal annotations indicating the start and end times of the single-speaker segments.

\subsection{Fine-grained Segmentation by VAD}\label{sec:vad}
Although the speaker diarization pipeline provides a rough segmentation of the raw speech data, some segments may still be too long to fit into memory. To address this, we use the Silero-VAD model\footnote{\url{https://github.com/snakers4/silero-vad}} to further split any segment longer than 30 seconds into smaller VAD segments. We then concatenate consecutive VAD segments from the same speaker into appropriately sized utterances, ensuring each utterance is between 3 and 30 seconds in length.

\subsection{ASR}\label{sec:asr}
The absence of text transcriptions limits the direct use of the \textit{Emilia} dataset for TTS tasks. To address this, we apply ASR techniques to transcribe the segmented speech data. Balancing speed and accuracy, we utilize the state-of-the-art multilingual ASR model, Whisper-Medium.\footnote{\url{https://huggingface.co/openai/whisper-medium}} To further enhance efficiency, we use WhisperX \cite{whipserx}, which is built on the faster-whisper\footnote{\url{https://github.com/SYSTRAN/faster-whisper}} backend and the CTranslate2\footnote{\url{https://github.com/OpenNMT/CTranslate2}} inference engine. This setup is up to four times faster than the official Whisper implementation while maintaining nearly the same accuracy. To avoid redundant processing, we bypass WhisperX’s VAD component by using the results from Sec.~\ref{sec:vad}. Additionally, we implement batched inference with the faster-whisper backend to transcribe speech data in parallel. These optimizations significantly improves the efficiency of the pipeline.

\subsection{Filtering}
In real-world scenarios, some noise may not be fully addressed by source separation, the Whisper model may have hallucinations, and some raw speech data may be of low quality~\cite{AutoPrep}. To ensure the quality of the resulting dataset, we apply the following filtering criteria.\footnote{These criteria can be adjusted for the specific needs of different use cases.}
First, we use the language identification results from the Whisper model in Sec.~\ref{sec:asr} and discard any speech data not predicted to be in our target languages (English, French, German, Chinese, Japanese, Korean) or with model language confidence below 80\%.
Secondly, we use the DNSMOS P.835 OVRL score~\cite{dnsmos835} to assess overall speech quality, retaining only speech data with a score higher than 3.0. Finally, for each raw speech sample, we calculate the average character duration across its corresponding segments. Segments with an average phone duration that falls outside 1.5 times the interquartile range (IQR) above the third quartile or below the first quartile are considered outliers, and the associated speech segments are discarded. After filtering, the resulting dataset is prepared for training the speech generation model.
\subsection{Performance Evaluation} \label{sec:pip_exp}
\begin{table*}[htbp]
    \centering
    \caption{Statistics of 600 hours in-the-wild speech data processed by Emilia-Pipe.}
    \label{tab:experimental_results}
    \begin{tabular}{cccccccc}
        \toprule
        \multirow{2}{*}{\textbf{Dataset}} & \multicolumn{3}{c}{\textbf{Duration (s)}} & \multicolumn{3}{c}{\textbf{DNSMOS P.835 OVRL}} & \multirow{2}{*}{\textbf{Total Duration (hours)}}\\
        \cmidrule(lr){2-4} \cmidrule(lr){5-7}
         & \textbf{min} & \textbf{max} & \textbf{avg \textpm~std} & \textbf{min} & \textbf{max} & \textbf{avg \textpm~std}  &\\
        \midrule
        Raw & \centering 9.22 & 18,056.98 & 1,572.53 \textpm~1,966.66 & 1.08 & 3.51 & 2.50 \textpm~0.62 & 598.87 (100.00\%)\\  
        Processed w/o Filtering & 1.00 & 30.00 & 7.18 \textpm~5.06 & 0.91 & 3.67 & 2.86 \textpm~0.51 & 340.54 (56.86\%) \\
        Processed & 3.00 & 30.00 & 8.98 \textpm~4.99 & 3.00 & 3.67 & 3.26 \textpm~0.14 & 176.22 (29.43\%) \\
        \bottomrule
    \end{tabular}
\end{table*}

To analyze Emilia-Pipe, we randomly sample a subset of raw speech data, approximately 600 hours, and use Emilia-Pipe to process this subset to evaluate its effectiveness and efficiency. 

The evaluation is conducted on an independent server with eight NVIDIA RTX 4090 GPUs. The whole processing time takes about 3.99 hours.
Table~\ref{tab:experimental_results} shows the processing results of Emilia-Pipe on this subset. The raw data has a wide range of audio durations from 9.22 to 18,056.98 seconds, with an average of 1,572.53 seconds and high variability. The DNSMOS P.835 OVRL scores range from 1.08 to 3.51, with an average of 2.50, indicating varied overall quality. After filtering, the total duration for the resulting data is further reduced to 176.22 hours, retaining 29.43\% of the raw speech data, and the average DNSMOS P.835 OVRL score significantly improves to 3.26 with minimal variability, indicating that Emilia-Pipe can effectively transform in-the-wild speech data into high-quality training data for speech generation. Besides, for processing this subset, Emilia-Pipe processes about 2.50 hours of data every one minute. This demonstrates that our processing method significantly exceeds real-time standards, making it ideal for preprocessing extensive speech data and scaling up the training dataset.

\section{The Emilia Dataset}
\label{dataset}
\subsection{Overview}

\begin{figure}[htbp]
    \centering
    \resizebox{0.3\textwidth}{!}{
        \begin{tikzpicture}
            \definecolor{color1}{HTML}{447cac}
            \definecolor{color2}{HTML}{88ce9b}
            \definecolor{color3}{HTML}{e3f79b}
            \definecolor{color4}{HTML}{fae28c}
            \definecolor{color5}{HTML}{f1874b}
            \definecolor{color6}{HTML}{c42d40}
            \pie[
                text=legend,
                radius=3,
                color={color1, color2, color3, color4, color5, color6},
                explode=0.1, 
                sum=auto, 
                before number=\phantom{0}, 
                after number=\% 
            ]{
                46.77/En: 46.8k hrs, 
                49.87/Zh: 49.9k hrs, 
                1.59/De: 1.6k hrs, 
                1.38/Fr: 1.4k hrs, 
                1.72/Ja: 1.7k hrs, 
                0.22/Ko: 0.2k hrs
            }
        \end{tikzpicture}
    }
    \caption{Duration statistics of the speech data by language.}
    \label{fig:dataset_stats}
\end{figure}
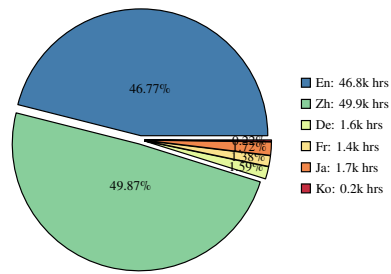

Using Emilia-Pipe, we construct the Emilia dataset from a vast collection of speech data sourced from diverse video platforms and podcasts on the Internet, covering various content categories such as talk shows, interviews, debates, sports commentary, and audiobooks. This variety ensures the dataset captures a wide array of real human speaking styles.
After processing, the initial version of the Emilia dataset includes a total of 101,654 hours of multilingual speech data in six different languages: English, French, German, Chinese, Japanese, and Korean. Fig~\ref{fig:dataset_stats} provides the duration statistics for each language in the dataset.

\subsection{Dataset Analysis}
In this subsection, we analyze the quality and diversity of the Emilia.

\subsubsection{Quality}
To evaluate quality, we compared Emilia with existing datasets using DNSMOS P.835 OVRL scores. This non-intrusive speech quality metric reflects the overall quality of the speech data and is highly correlated with human ratings~\cite{dnsmos835}. Table~\ref{tab:dnsmos_results} presents the speech quality comparison between Emilia and several existing datasets.
Emilia achieves a DNSMOS P.835 OVRL score of 3.26, ranking third among all datasets. The results indicate that, despite being sourced from raw speech data in the wild, after preprocessing, the speech quality of the Emilia dataset is comparable to existing datasets sourced from studio recordings or audiobooks and outperforms all existing datasets sourced from in-the-wild speech data.

\begin{table}[htbp]
\centering
\caption{Quality comparison between Emilia and nine existing datasets. The scores for LJSpeech, AutoPrepWild, Aishell-3, and LibriTTS are derived from~\cite{AutoPrep}. The score for Libri-Light is computed from its official "small" subset, and the score for WenetSpeech4TTS is computed from its official "basic" subset. The scores for MLS and Emilia are computed from a randomly sampled 600-hour subset.}
\label{tab:dnsmos_results}
\begin{tabular}{cc}
\toprule
\textbf{Dataset} & \textbf{DNSMOS P.835 OVRL} \\
\midrule
LJSpeech~\cite{ljspeech} & 3.30 ± 0.17 \\
AutoPrepWild~\cite{AutoPrep} & 3.24 ± 0.21 \\
VCTK \cite{vctk} & 3.20 ± 0.18\\
Aishell-3~\cite{aishell3} & 3.15 ± 0.17 \\
LibriTTS~\cite{libritts} & 3.25 ± 0.19 \\
GigaSpeech~\cite{gigaspeech} & 2.52 ± 0.54 \\
WenetSpeech4TTS~\cite{wenetspeech4tts} & 3.18 ± 0.22 \\
MLS \cite{mls}~& \textbf{3.33 ± 0.19} \\
Libri-Light~\cite{librilight} & 3.25 ± 0.26 \\
\midrule
Emilia & 3.26 ± 0.14 \\
\bottomrule
\end{tabular}
\end{table}

\subsubsection{Diversity}
The Emilia dataset comprises a collection of speech data from a wide range of video platforms and podcasts, capturing diverse speaking styles of real human speech. To quantify this diversity, we conducted analyses on both the acoustic and semantic feature space, comparing it with the MLS dataset, which is derived from audiobooks and widely used for training speech generation models.

\begin{figure}[htbp]
    \centering
    \subfloat[Acoustic diversity]{\includegraphics[width=.49\columnwidth]{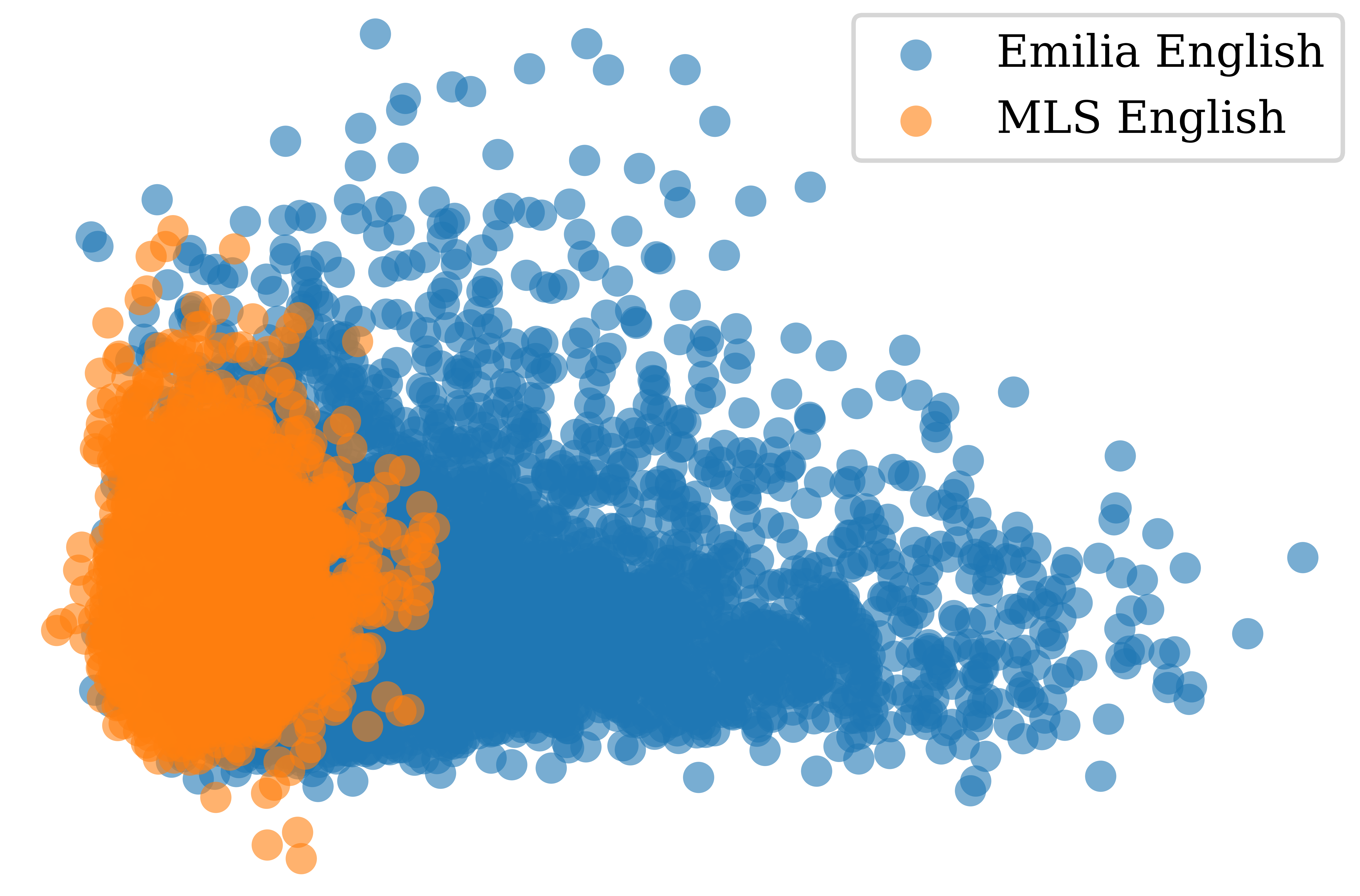}} \hspace{2pt}
    \subfloat[Semantic diversity]{\includegraphics[width=.49\columnwidth]{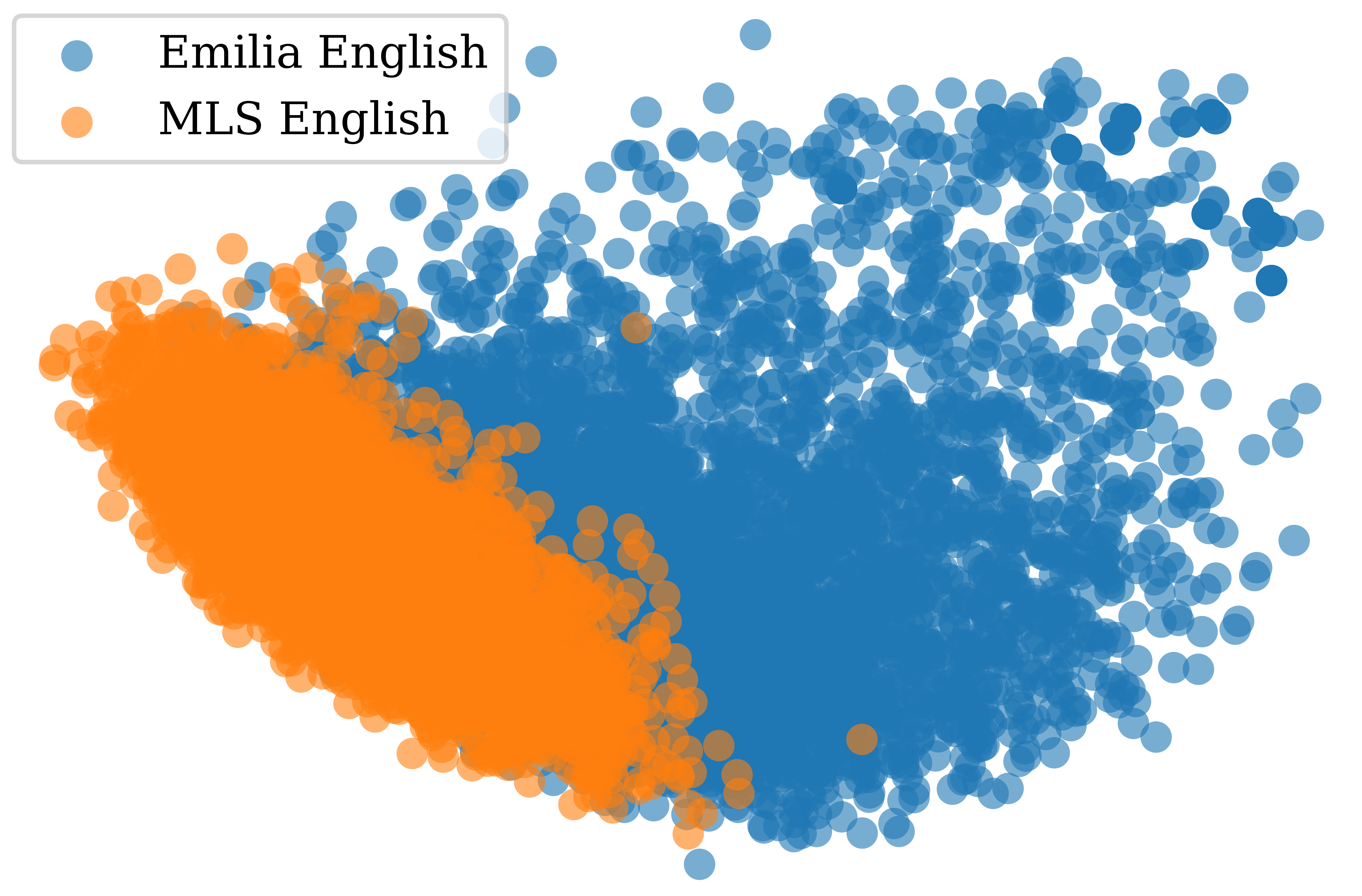}}
    \caption{A comparison of acoustic and semantic diversities between Emilia and MLS datasets.}
    \label{fig:combined}
\end{figure}

Specifically, we randomly select 5,000 samples each from the English subset of MLS and Emilia. To analyze the diversity of acoustic features, we leverage a pre-trained WavLM model\footnote{\url{https://huggingface.co/microsoft/wavlm-base-plus}} to extract acoustic representations, capturing a variety of acoustic characteristics such as speaker, emotion, and prosody~\cite{chen2022wavlm}. We then apply the PCA algorithm to reduce the dimensionality of these representations to two.
As shown in Fig.~\ref{fig:combined} (a), the Emilia dataset exhibits a broader dispersion, contrasting with MLS, which shows a more compact clustering. The more scattered pattern highlights the Emilia dataset as encompassing a richer acoustic characteristic coverage compared to the MLS dataset derived from audiobooks. 

For the semantic diversity analysis, we employ a pre-trained Sentence-BERT model\footnote{\url{https://github.com/UKPLab/sentence-transformers}} to generate text representations for the transcripts of each speech data. Consequently, each speech data is represented as a 768-dimensional vector based on its textual content, providing a comprehensive approximation of its semantic \cite{sentencebert}. Similar to the analysis above, we reduced the dimension of the semantic features to two. As shown in Fig.~\ref{fig:combined} (b), the scatter of textual features indicates that the Emilia dataset covers a wide range of textual content, validating the significant diversity in Emilia's semantic coverage.

\begin{table*}[htbp]
\centering
\caption{Objective and subjective evaluation of TTS models using Emilia and MLS on LibriSpeech-Test and Emilia-Test evaluation sets.}
\label{tab:combined_evaluation}
\begin{tabular}{ccccccccccccc}
\toprule
\multirow{2}{*}{\textbf{Model}} & \multirow{2}{*}{\textbf{Train Set}} & \multicolumn{5}{c}{\textbf{LibriSpeech-Test}} & \multicolumn{5}{c}{\textbf{Emilia-Test}} \\ 
\cmidrule(lr){3-7} \cmidrule(lr){8-12}
 &  & \textbf{WER $\downarrow$} & \textbf{SIM-O $\uparrow$} & \textbf{FSD $\downarrow$} & \textbf{CMOS $\uparrow$} & \textbf{SMOS $\uparrow$} & \textbf{WER $\downarrow$} & \textbf{SIM-O $\uparrow$} & \textbf{FSD $\downarrow$} & \textbf{CMOS $\uparrow$} & \textbf{SMOS $\uparrow$} \\ 
\midrule
\multirow{2}{*}{AR} 
& MLS & 
8.9\% & 0.612 & 49.11 &  $-0.36$ & $3.13$ & 
7.7\% & 0.587 & 20.76 &  $0.09$ & $3.71$ \\ 
& Emilia & 
8.4\% & 0.577 & 24.73 &  $-0.19$ & $3.28$ & 
6.6\% & 0.618 & 12.73 & $0.19$ & $3.73$ \\ 
\midrule
\multirow{2}{*}{NAR} 
& MLS & 
6.1\% & 0.625 & 16.83 & $0.36$ & $3.62$ & 
8.2\% & 0.528 & 15.94 & $0.28$ & $3.61$ \\ 
& Emilia 
& 7.2\% & 0.585 & 23.24 & $0.42$ & $3.77$ &  
7.4\% & 0.601 & 14.07 & $0.28$ & $3.76$ \\ 
\bottomrule
\end{tabular}
\end{table*}

\section{Experiments}
In this section, we evaluate the effectiveness of the Emilia dataset in TTS task through two experiments: the English-only experiment and the multilingual experiment. In the English-only experiment, we compare the performance of TTS models trained with the English subset of the Emilia dataset to their counterparts trained with the English subset of the MLS dataset. In the multilingual experiment, we train the models with the full Emilia dataset, which comprises 101,654 hours of speech data in six languages, and evaluate their multilingual TTS performance.

\subsection{Experimental Setups}
\subsubsection{Baselines}
We implement two TTS models as baselines: AR+SoundStorm~\cite{borsos2023soundstorm} and VoiceBox~\cite{voicebox}. AR+SoundStorm uses an autoregressive (AR) model to predict speech semantic tokens from text \cite{borsos2023audiolm}, followed by SoundStorm~\cite{borsos2023soundstorm} to generate acoustic tokens in a non-autoregressive manner. VoiceBox is a fully non-autoregressive speech synthesis model that utilizes the flow-matching method and transformer architecture for speech generation, learning the distribution of the mel-spectrogram based on text input and speech context. For simplicity, we refer to AR+SoundStorm and VoiceBox as AR and NAR, respectively. 

Both baselines are trained using four NVIDIA A100 GPUs for over 800k steps to ensure convergence. The models are optimized with the AdamW optimizer at a learning rate of 5e-5, including 5k warmup steps, followed by a cosine annealing learning rate schedule.

\subsubsection{Evaluation Metrics}
To evaluate the baselines, we conduct both objective and subjective evaluations.

For the objective evaluation, we consider the following aspects: 
(1) Intelligibility: Measured by the Word Error Rate (WER) of the synthesized speech's transcription compared to the input text. For LibriSpeech-Test, we use a finetuned HuBERT-Large ASR model.\footnote{\url{https://huggingface.co/facebook/hubert-large-ls960-ft}} For other testsets, we use the Whisper-Medium model.\footnote{\url{https://huggingface.co/openai/whisper-medium}} 
(2) Coherence: Assessed by speaker similarity between generated speech and the speech prompt using the WavLM-TDCNN speaker embedding model.\footnote{\url{https://github.com/microsoft/UniSpeech/tree/main/downstreams/speaker_verification}}
We report similarity to the original speech prompt (SIM-O). 
(3) Naturalness: Evaluated using the Fréchet Speech Distance (FSD), which measures the similarity between the distributions of generated and real samples in a feature space. A lower FSD indicates higher speech quality and diversity~\cite{voicebox}. We adapt the metric for speech by using the emotion2vec features.\footnote{\url{https://github.com/ddlBoJack/emotion2vec/tree/main}} 

For the subjective evaluation, we randomly select eight samples each from LibriSpeech-Test and Emilia-Test. Twelve proficient English speakers served as judges. we use SMOS (Similarity Mean Opinion Score) to evaluate the speaker similarity of the speech to the original speech prompt. The SMOS scale ranges from 1 to 5, with increments of 0.5 points. CMOS (Comparative Mean Opinion Score) is used to evaluate the comparative naturalness of the synthesized speech against a given speech prompt. The CMOS scale ranges from -3 (indicating the synthesized speech is much worse than the speech prompt) to 3 (indicating the synthesized speech is much better than the speech prompt), with intervals of 1. 

\subsection{Emilia English versus MLS English}
The experiment evaluates the effectiveness of the proposed Emilia dataset by comparing the performance of models trained on the English subsets of both Emilia and the MLS dataset, a high-quality dataset derived from audiobooks. The total duration of the Emilia dataset is 46k hours, while the MLS dataset comprises 44.5k hours. The size of the datasets can be considered roughly equivalent. To thoroughly evaluate the models, we utilize the LibriSpeech-Test evaluation set, containing 1,200 speech samples in formal reading styles akin to those of MLS, and the Emilia-Test evaluation set, which includes 600 speech samples in diverse spontaneous speaking styles. Both test sets contain speech data that are unseen by the baselines.

Table~\ref{tab:combined_evaluation} presents the results of both objective and subjective evaluations for the Emilia and MLS datasets on the LibriSpeech-Test and Emilia-Test. From these results, we observe that models trained on both the Emilia and MLS datasets demonstrate similar levels of speaker similarity (measured by SIM-O and SMOS) and intelligibility (measured by WER). This suggests that the Emilia dataset, despite being sourced from raw speech data in the wild, is as effective as high-quality datasets derived from audiobooks after processing with our proposed Emilia-Pipe. The NAR model trained with both datasets demonstrates a similar level of naturalness. In contrast, the AR model shows significant improvement in FSD and CMOS on the Emilia-Test, which contains speech prompts in diverse spontaneous speaking styles. These results may indicate that AR TTS models benefit more from speech with diverse speaking styles compared to NAR models.
 
\subsection{Multilingual TTS}

\begin{table}[hbtp]
\centering
\caption{Objective evaluation of TTS models trained on Emilia on six languages.}
\label{tab:multilingual}
\begin{tabular}{cccccc}
\toprule
\textbf{Language} & \textbf{Model} & \textbf{WER $\downarrow$} & \textbf{SIM-O $\uparrow$} & \textbf{FSD $\downarrow$}\\ 
\midrule
\multirow{2}{*}{En} & AR & 6.2\% & 0.614 & 14.82 \\  
                    & NAR  & 5.8\% & 0.581& 14.47 \\ 
\midrule
\multirow{2}{*}{Zh} & AR        & 4.1\% & 0.564 & 35.59   \\  
                    & NAR  &4.7\% &0.560& 48.95  \\  
\midrule 
\multirow{2}{*}{De} & AR        & 6.8\% & 0.680 & 32.72   \\  
                    & NAR  &13.3\% & 0.633&44.68 \\
\midrule
\multirow{2}{*}{Fr} & AR        & 8.2\% & 0.589 & 42.92  \\  
                    & NAR  & 17.5\%&0.550& 48.82& \\ 
\midrule
\multirow{2}{*}{Ja} & AR & 3.6\% & 0.625 & 49.42  \\  
                    & NAR  & 10.8\%& 0.562 &49.39   \\  
\midrule
\multirow{2}{*}{Ko} & AR  & 10.9\% & 0.681 & 47.93 \\  
                    & NAR  &15.2\% &0.608&63.00  \\
\bottomrule
\end{tabular}
\end{table}

Next, we conduct experiments to evaluate the performance of the multilingual TTS capability of the baselines trained on the full Emilia datasets, encompassing six languages. The test set for En is Emilia-Test. The test set for Zh is from Aishell-3. The test sets for De, Fr, Ja, and Ko are from Common Voice. Each test set contains at least 500 reference samples. Due to space limitations, we only present the objective evaluation results in Table~\ref{tab:multilingual}, omitting the subjective evaluation results. The findings confirm that both models exhibit strong zero-shot multilingual TTS performance, underscoring the multilingual effectiveness of the Emilia dataset.

\section{Conclusions}
In conclusion, this paper introduces Emilia, an extensive, multilingual, and diverse dataset for speech generation, along with Emilia-Pipe, an open-source preprocessing pipeline that can effectively and efficiently transforms raw speech into high-quality training data. The initial version of the Emilia dataset includes over 101k hours of speech data in six languages, featuring a wide variety of in-the-wild real-world speech. Both objective and subjective evaluations demonstrate that training on this diverse dataset can significantly advance speech generation, enabling models to produce more natural, spontaneous, and human-like speech. We have open-sourced both the \textit{Emilia} dataset and the \textit{Emilia-Pipe} pipeline, inviting the research community to collaborate on large-scale, spontaneous speech generation initiatives and further drive innovation in this field.

\bibliographystyle{IEEEbib}
\bibliography{main}

\end{document}